\documentclass[a4paper]{spie}
\usepackage[]{graphicx}

\newcommand{\wmap}{{\scshape\footnotesize{WMAP~}}}

\newcommand{\be}{\begin{equation}}
\newcommand{\ee}{\end{equation}}

\def\ltsima{$\; \buildrel < \over \sim \;$}
\def\simlt{\lower.5ex\hbox{\ltsima}}
\def\gtsima{$\; \buildrel > \over \sim \;$}
\def\simgt{\lower.5ex\hbox{\gtsima}}

\title{The cosmic microwave background: \\ observing directly the early universe}

\author{Paolo de Bernardis\supit{a} and Silvia Masi\supit{a}
\skiplinehalf \supit{a}Dipartimento di Fisica, Sapienza
Universit\'a di Roma, P.le A. Moro 2 00185 Roma, Italy }

\authorinfo{Further author information: (Send correspondence to PdB)\\PdB.: E-mail: paolo.debernardis@roma1.infn.it,
 Telephone: 39 06 49914 271\\  SM: E-mail: silvia.masi@roma1.infn.it, Telephone: +39 06 49914 271}

\begin{document}
\maketitle


\begin{abstract}
The Cosmic Microwave Background (CMB) is a relict of the early
universe. Its perfect 2.725K blackbody spectrum demonstrates that
the universe underwent a hot, ionized early phase; its anisotropy
(about 80 $\mu$K rms) provides strong evidence for the presence of
photon-matter oscillations in the primeval plasma, shaping the
initial phase of the formation of structures; its polarization
state (about 3 $\mu$K rms), and in particular its rotational
component (less than 0.1 $\mu$K rms) might allow to study the
inflation process in the very early universe, and the physics of
extremely high energies, impossible to reach with accelerators.
The CMB is observed by means of microwave and mm-wave telescopes,
and its measurements drove the development of ultra-sensitive
bolometric detectors, sophisticated modulators, and advanced
cryogenic and space technologies. Here we focus on the new
frontiers of CMB research: the precision measurements of its
linear polarization state, at large and intermediate angular
scales, and the measurement of the inverse-Compton effect of CMB
photons crossing clusters of Galaxies. In this framework, we will
describe the formidable experimental challenges faced by
ground-based, near-space and space experiments, using large arrays
of detectors. We will show that sensitivity and mapping speed
improvement obtained with these arrays must be accompanied by a
corresponding reduction of systematic effects (especially for CMB
polarimeters), and by improved knowledge of foreground emission,
to fully exploit the huge scientific potential of these missions.
\end{abstract}



\section{INTRODUCTION}
\label{sec:intro}  

We live in an expanding universe, cooling down from a state of
extremely high density and temperature, the big bang. In our
universe the ratio between the density of photons (the photons of
the cosmic microwave background) and the density of baryons is of
the order of 10$^9$: this abundance of photons dominated the
dynamics of the Universe in the initial phase (first 50000 years).
During the first 380000 years the universe was ionized and opaque
to radiation, due to the tight coupling between photons and
charged baryons. Radiation thermalized in this primeval fireball,
producing a blackbody spectrum. When the universe cooled down
below 3000K, neutral atoms formed (recombination), and radiation
decoupled from matter, traveling basically without any further
interaction all the way to our telescopes. Due to the expansion of
the universe, the wavelengths of photons expand (by the same
amount all lengths expanded, a factor of 1100). What was a glowing
3000K blackbody 380000 years after the big bang, has been
redshifted to millimeter waves, and is now observable as a faint
background of microwaves. This is the cosmic microwave background,
which has been observed as a 2.725K blackbody filling the present
universe \cite{Math99}.

The CMB is remarkably isotropic. However, it is widely believed
that the large scale structure of the universe observed today (see
e.g. \cite{Padm93}) derives from the growth of initial density
seeds, already visible as small anisotropies in the maps of the
Cosmic Microwave Background. This scenario works only if there is
dark (i.e. not interacting electromagnetically) matter, already
clumped at the epoch of CMB decoupling, gravitationally inducing
anisotropy in the CMB. There are three physical processes
converting the density perturbations present at recombination into
{\it observable } CMB temperature fluctuations $\Delta T / T$.
They are: the photon density fluctuations $\delta_\gamma$, which
can be related to the matter density fluctuations $\Delta \rho$
once a specific class of perturbations is specified; the
gravitational redshift of photons scattered in an over-density or
an under-density  with gravitational potential difference $\Delta
\phi_r$; the Doppler effect produced by the proper motion with
velocity $v$ of the electrons scattering the CMB photons. In
formulas:
\begin{equation} \label{dtt}
\frac{\Delta T}{T}(\vec{n}) \approx \frac{1}{4}\delta_{\gamma r}+
\frac{1}{3} {\Delta \phi_r \over c^2} - \vec{n} \cdot {\vec{v_r}
\over c}
\end{equation}
where $\vec{n}$ is the line of sight vector and the subscript $r$
labels quantities at recombination.

\begin{figure}
\begin{center}
\begin{tabular}{c}
\includegraphics[height=8.4cm, width=7.5cm]{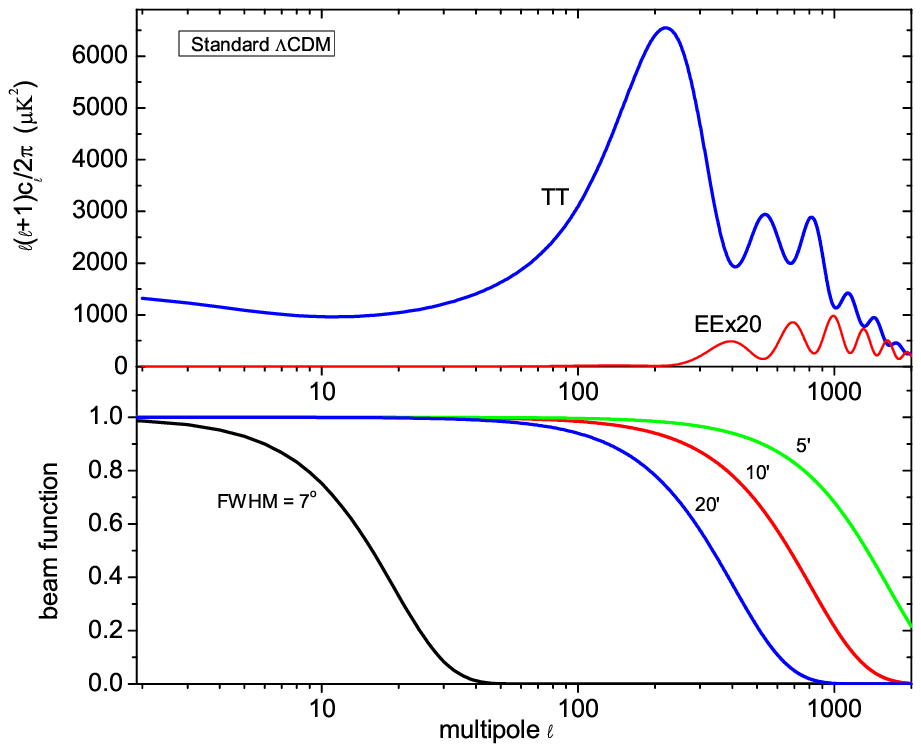}
\includegraphics[height=8cm, width=9.5cm]{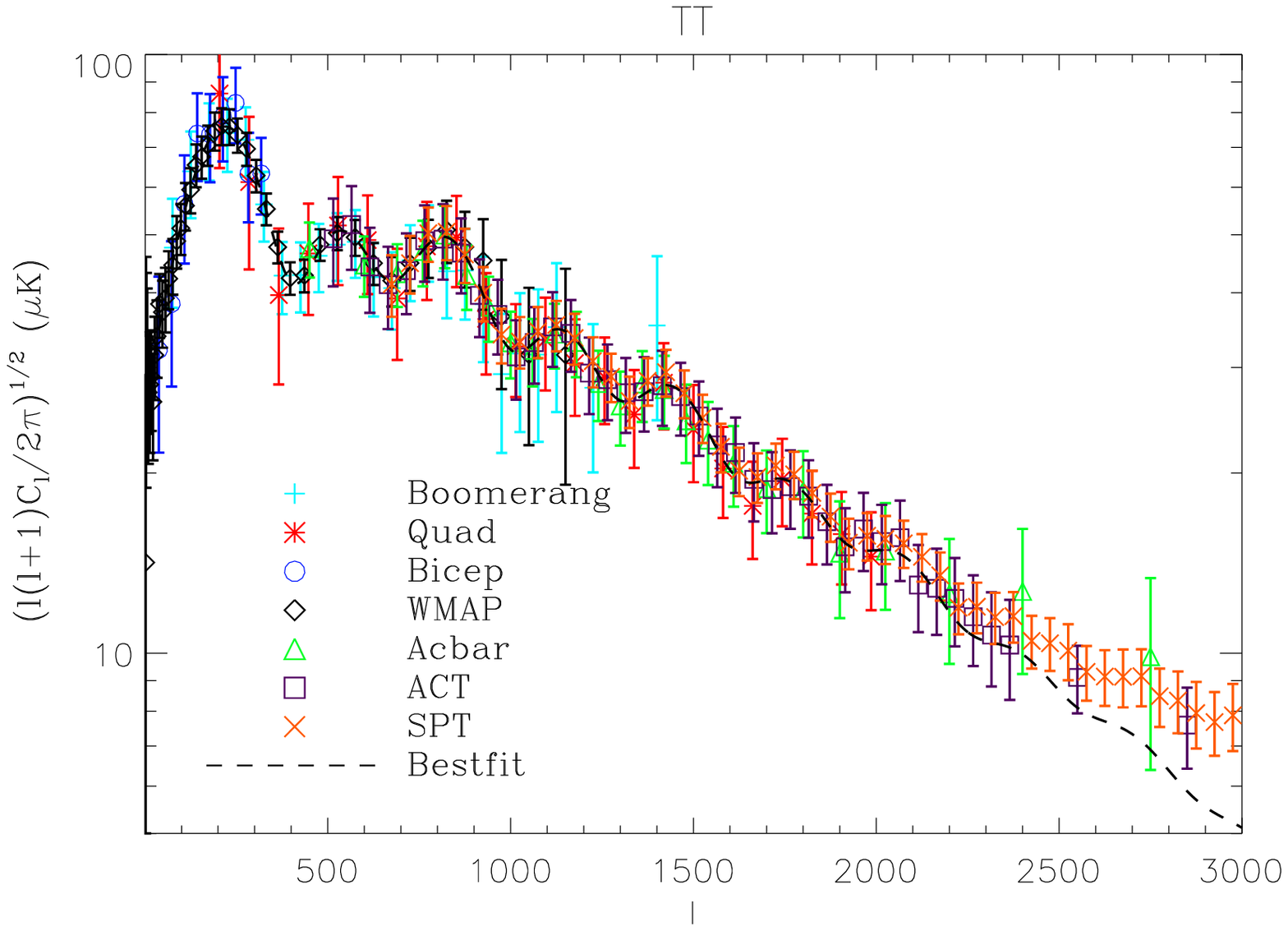}
\end{tabular}
\end{center}
\caption[fig:spettri] {\label{fig:spettri}{\bf Left-Top:} Angular
power spectrum for CMB anisotropy (TT) and for EE polarization.
The latter has been amplified 20 times to make it visible in the
same plot of TT. The angular scale $\gamma$ corresponding to
multipole $\ell$ is approximately $\gamma(^o) = 180/\ell$. {\bf
Left-Bottom:} Filter functions of CMB telescopes with different
angular resolutions. A FWHM smaller than 1$^o$ is needed to be
sensitive to the "acoustic peaks" due to photon-baryon
oscillations in the early universe. The curves are labeled with
the beam FWHM. Differential instruments will not be sensitive to
multipoles $\ell < 180/\alpha(^o)$ where $\alpha$ is the angular
separation of the beam switch; experiments scanning a limited sky
region with angular size $\theta$ cannot be sensitive to
multipoles with $\ell < 180/\theta(^o)$ . {\bf Right:} Selected
recent measurements of the angular power spectrum of CMB
anisotropy. }
\end{figure}

Our description of fluctuations with respect to the FRW isotropic
and homogeneous metric is totally statistical. So we are not able
to forecast the map $\Delta T / T$ as a function of $(\theta,
\phi)$, but we are able to predict its statistical properties. If
the fluctuations are random and Gaussian, all the information
encoded in the image is contained in the angular power spectrum of
the map, detailing the contributions  of the different angular
scales to the fluctuations in the map. In other words, the power
spectrum of the image of the CMB details the relative abundance of
the spots with different angular scales.

If we expand the temperature of the CMB in spherical harmonics, we
have
\begin{equation} \label{multi}
{\Delta T \over T} = \sum a_{\ell,m}^T Y_\ell^m (\theta, \phi)
\end{equation}
the power spectrum of the CMB temperature anisotropy is defined as
\begin{equation}
c_\ell^{TT} = \langle TT \rangle = \langle a_{\ell,m}^T
a_{\ell,m}^{T,*} \rangle
\end{equation}
with no dependence on $m$ since there are no preferred directions.
Since we have only a statistical description of the observable,
the precision with which the theory can be compared to measurement
is limited both by experimental errors {\sl and} by the
statistical uncertainty in the theory itself. Each observable has
an associated cosmic and sampling variance, which depends on how
many independent samples can be observed in the sky. In the case
of the $c_\ell$s, their distribution is a $\chi^2$ with $2\ell+1$
degrees of freedom, which means that low multipoles have a larger
intrinsic variance than high multipoles (see e.g. \cite{Whit93}).

Theory predicts the angular power spectrum of CMB anisotropy with
remarkable detail, given a model for the generation of density
fluctuations in the Universe, and a set of parameters describing
the background cosmology. Assuming scale-invariant initial density
fluctuations, the main features of the power spectrum
$c_\ell^{TT}$ are a $1/\left[\ell(\ell + 1) \right]$ trend at low
multipoles, produced by the Sachs-Wolfe effect\cite{Sach67}
(second term in equation \ref{dtt}); a sequence of peaks and dips
at multipoles above $\ell=100$, produced by acoustic fluctuations
in the primeval plasma of photons and
baryons\cite{Suny70,Peeb70,Hu96}, and a damping tail at high
multipoles, due to the finite depth of the recombination and
free-streaming effects\cite{Hu97a}. Detailed models and codes are
available to compute the angular power spectrum of the CMB image
(see e.g. \cite{Hu02,Lewi00}). The power spectrum $c_\ell^{TT}$
derived from the current best-fit cosmological model is plotted in
fig.\ref{fig:spettri} (top panel).

\begin{figure}
\begin{center}
\begin{tabular}{c}
\includegraphics[height=6.5cm, angle=90]{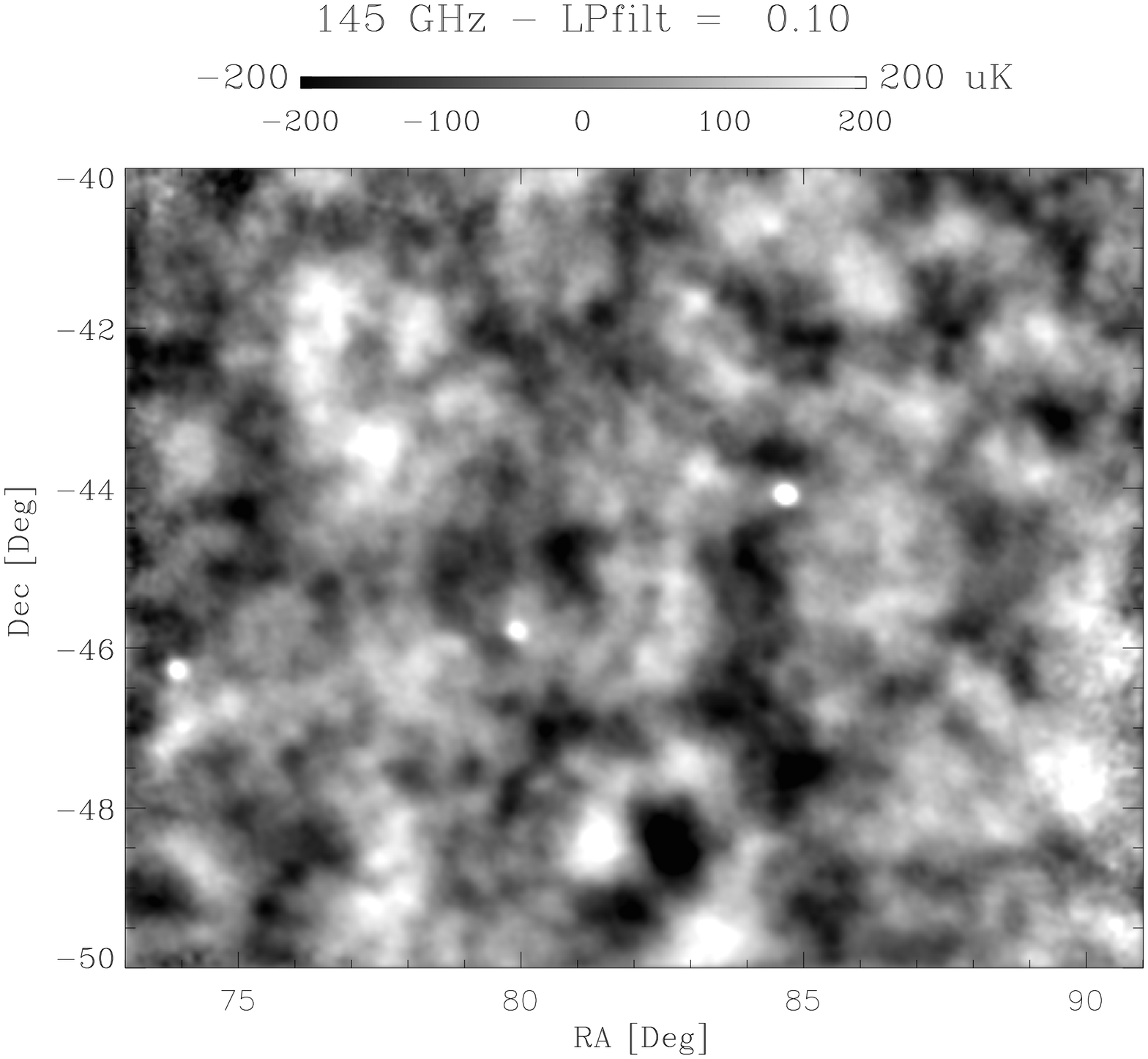}
\includegraphics[height=10.5cm, angle=90]{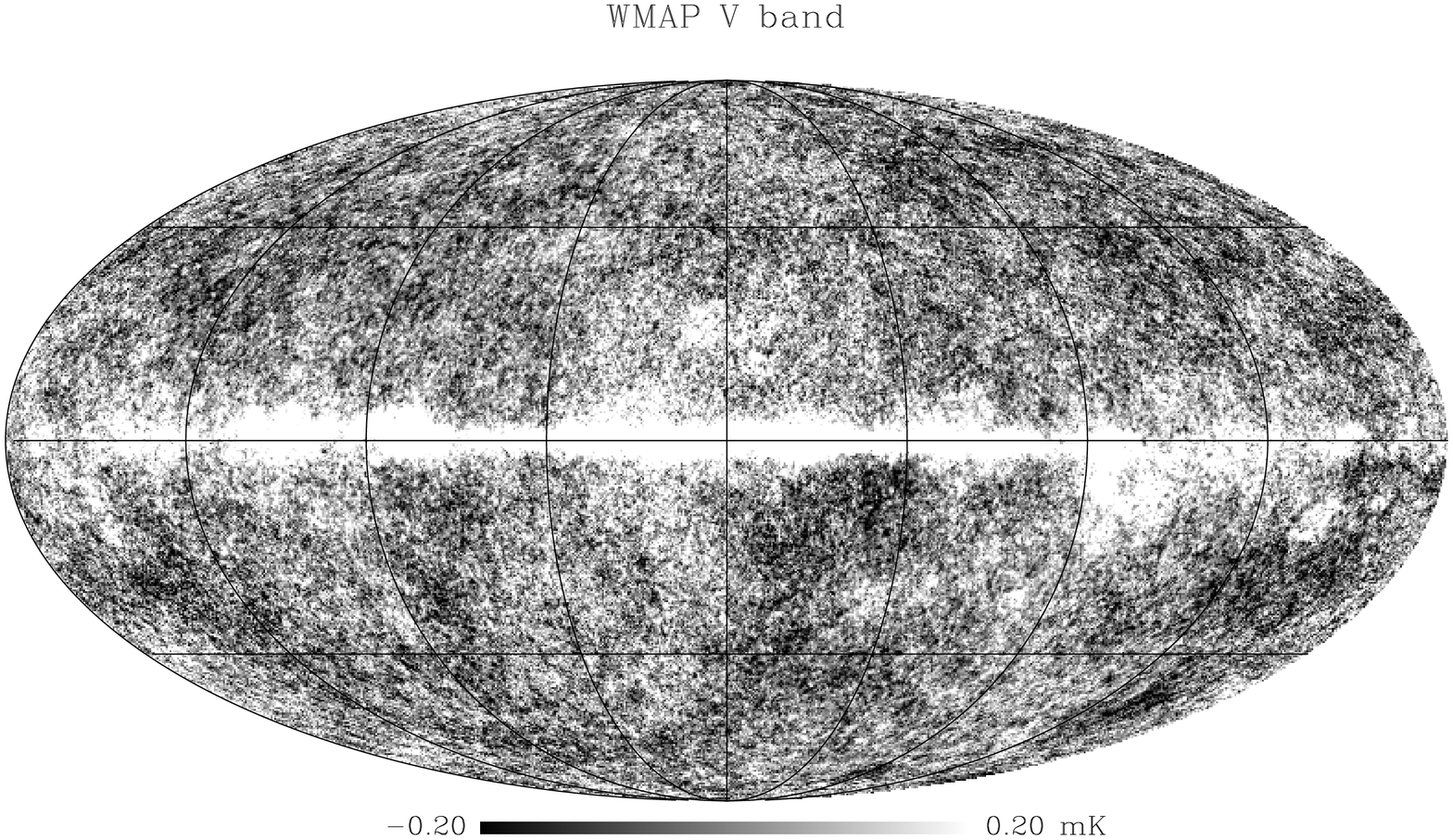}
\end{tabular}
\end{center}
\caption[fig:mappa] {\label{fig:mappa} {\bf Left:} The first map
of the CMB with angular resolution and signal to noise ratio
sufficient to resolve degree-sized causal horizons in the early
universe was obtained by the BOOMERanG experiment, at 145 GHz,
using an off-axis telescope flown on a stratospheric
balloon\cite{debe00}. The structures visible in the map are CMB
anisotropies, while the contamination from local foregrounds and
from instrument noise are both negligible. {\bf Right:} The WMAP
satellite has mapped the whole sky, confirming the ubiquitous
presence of causal horizons, and allowing a precise determination
of the power spectrum of CMB anisotropy and of the cosmological
parameters \cite{Benn03,Lars11}.}
\end{figure}

High signal-to-noise maps of the CMB have been obtained since year
2000 \cite{debe00} (see fig.\ref{fig:mappa}).
From such maps, the power spectrum of CMB anisotropy is now
measured quite well (see e.g.
\cite{Wrig91,debe00,Lee01,debe02,Halv02,Ruhl03,Grai03,
Jone06,Kuo07,Hins03,Hins07,Nolt09,Kuo09,Fowl10,Lars11,Das11} and
fig.\ref{fig:spettri}, right panel); moreover, higher order
statistics are now being measured with the accuracy required to
constrain cosmological parameters (see e.g. \cite{Das11b}).
Despite of the very small signals, the measurements from
independent experiments, using diverse experimental techniques,
are remarkably consistent. Moreover, an adiabatic inflationary
model, with cold dark matter and a cosmological constant, fits
very well the measured data (see e.g. \cite{debe94, Bond98,
Bond00, Dode00,
Tegm00a,Tegm00b,Brid01,Dous01,Lang01,Jaff01,Lewi02, Nett02,
Ruhl03, Sper03, Benn03, Tegm04, Sper07, Koma09, Koma11}).

The CMB is expected to be slightly polarized, since most of the
CMB photons undergo a last Thomson scattering at recombination,
and the radiation distribution around the scattering centers is
slightly anisotropic. Any quadrupole anisotropy in the incoming
distribution produces linear polarization in the scattered
radiation. The main term of the local anisotropy due to density
(scalar) fluctuations is dipole, while the quadrupole term is much
smaller.  For this reason the expected polarization is quite weak
(\cite{Rees68,Kais83,Hu97,Kami97}). The polarization field can be
expanded into a curl-free component (E-modes) and a curl component
(B-modes). Six auto and cross power spectra can be obtained from
these components: $\langle TT \rangle$, $\langle TE \rangle$, $
\langle EE \rangle$, $\langle BB \rangle$, $\langle TB \rangle$,
and $\langle EB \rangle$. For example

\begin{equation}
c_\ell^{TE} = \langle TE \rangle = \langle a_{\ell,m}^T
a_{\ell,m}^{E,*} \rangle
\end{equation}

\noindent where the $a_{\ell,m}^E$ and $a_{\ell,m}^B$ decompose
the map of the Stokes parameters $Q$ and $U$ of linear
polarization in spin-2 spherical harmonics:

\begin{equation}
(Q \pm i U)(\theta, \phi) = \sum_{\ell m} \left(a_{\ell m}^E \mp i
a_{\ell m}^B \right)  ~_{\pm 2}Y_{\ell m}(\theta, \phi)
\end{equation}.

Due to the parity properties of these components, standard
cosmological models have $\langle TB \rangle=0$ and $\langle EB
\rangle=0$. Linear scalar (density) perturbations can only produce
E-modes of polarization (see e.g. \cite{Selj97}). In the
concordance model, $\langle EE \rangle \sim 0.01 \langle TT
\rangle$, making $\langle EE \rangle$ a very difficult observable
to measure. The power spectrum $c_\ell^{EE}$ derived from the
current best-fit cosmological model is plotted in
fig.\ref{fig:spettri} (top-left panel).

Tensor perturbations (gravitational waves) produce both E-modes
and B-modes. If inflation  happened (see e.g.
\cite{Mukh81,Guth82,Lind83,Kolb90}), it produced a weak background
of gravitational waves. The resulting level of the B-modes depends
on the energy scale of inflation, but is in general very weak (see
e.g. \cite{Cope93,Turn93}). Alternative scenarios, like the cyclic
model\cite{Stei02}, do not produce B-modes at all\cite{Boyl04}.

There is a strong interest in measuring CMB polarization, and in
particular the B-modes, because their detection would represent
the final confirmation of the inflation hypothesis, and their
level would constrain the energy-scale of the inflation process,
which, we know, happened at extremely high energies (which cannot
be investigated on earth laboratories \cite{Lidd94}).

For long time attempts to measure CMB polarization resulted in
upper limits (see e.g.
\cite{Cade78,Nano79,Lubi81,Masi86,Part88,Woll97}). The possibility
of detecting the $\langle BB \rangle$ signature of the
inflationary gravity waves background renewed the interest in
these measurements
(\cite{Keat01,Subr00,Hedm02,Picc02,Dela02,Masi02,Vill02,Kova02,John03,Keat03a,Keat03b,
Kogu03,Fare04,Cort04,Cart05}). The first statistically significant
detections of CMB polarization have been reported by the coherent
radiometer experiments DASI\cite{Leit05}, CAPMAP\cite{Bark05},
CBI\cite{Read04}, WMAP for both $\langle TE \rangle$\cite{Kogu03}
and $\langle EE \rangle$\cite{Page07}, and by the bolometric
instrument BOOMERanG-03\cite{Masi06,Piac06,Mont06}. The quality of
CMB polarization measurements has improved steadily with the
introduction of instruments with detectors arrays, like
QUAD\cite{Quad08,Quad09a,Quad09b}, BICEP\cite{Chia10}, and
QUIET\cite{Bisc11}.

\begin{figure}
\begin{center}
\begin{tabular}{c}
\includegraphics[height=6.5cm, width=5.5cm]{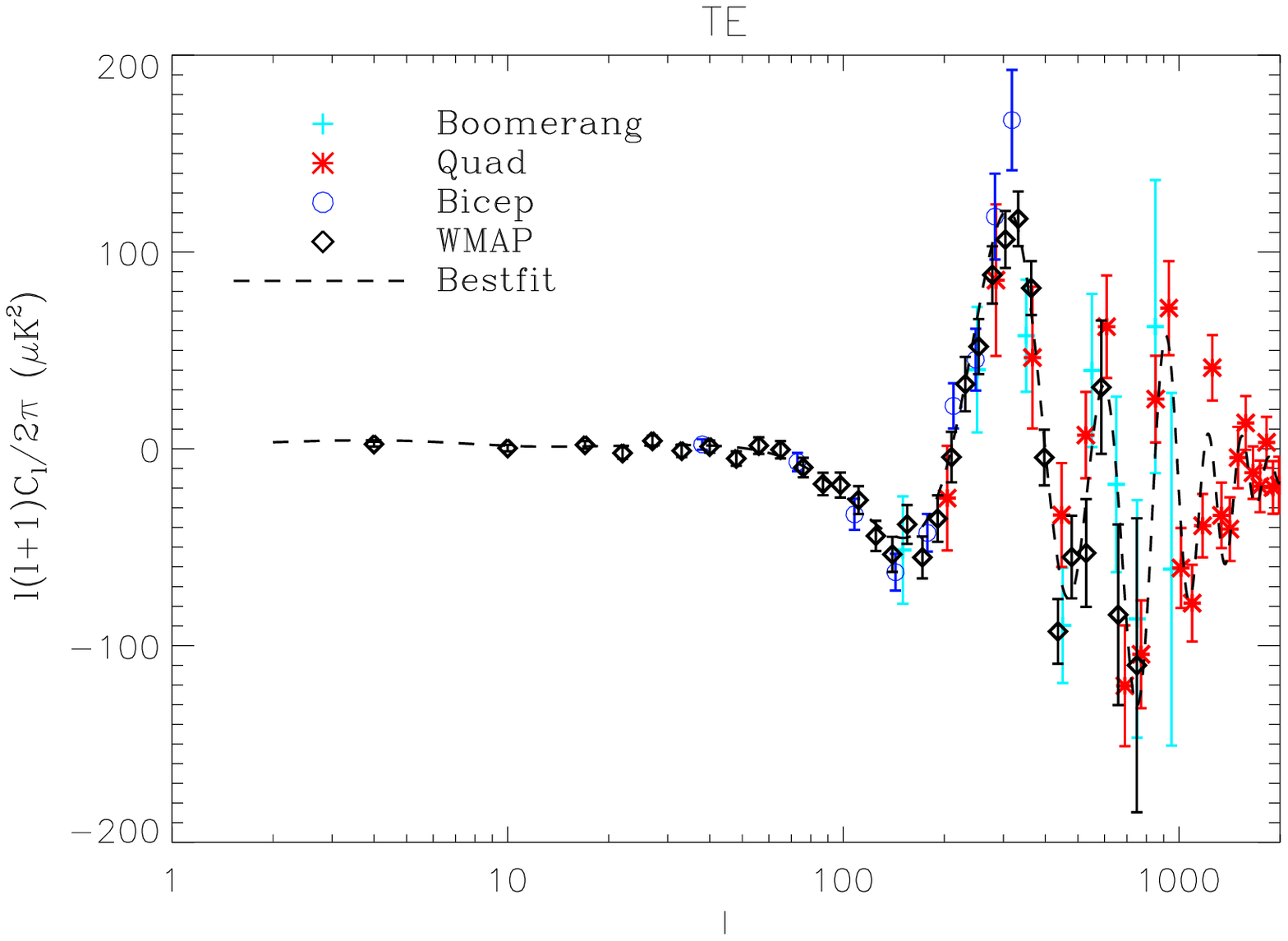}
\includegraphics[height=6.5cm, width=5.5cm]{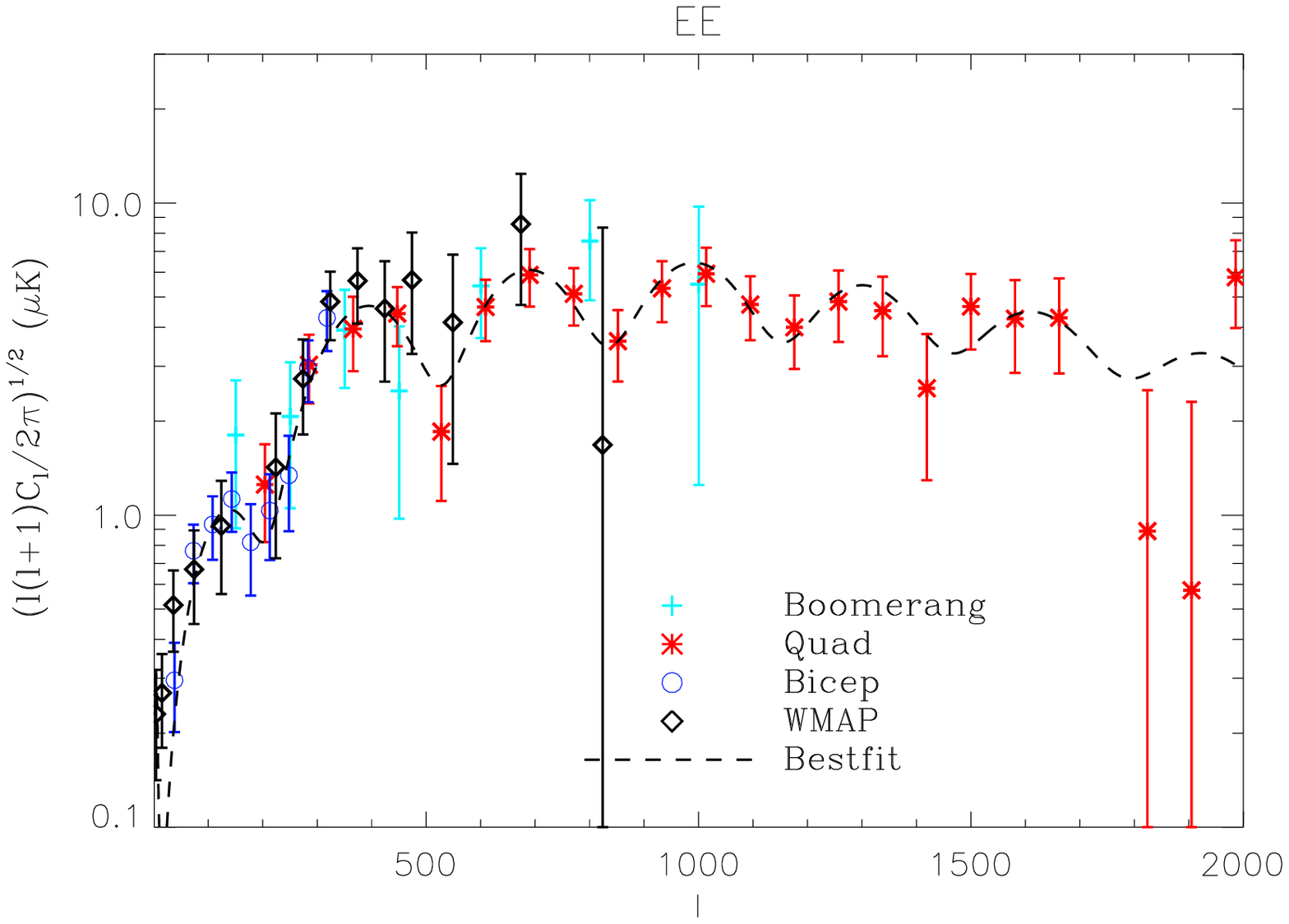}
\includegraphics[height=6.5cm, width=5.5cm]{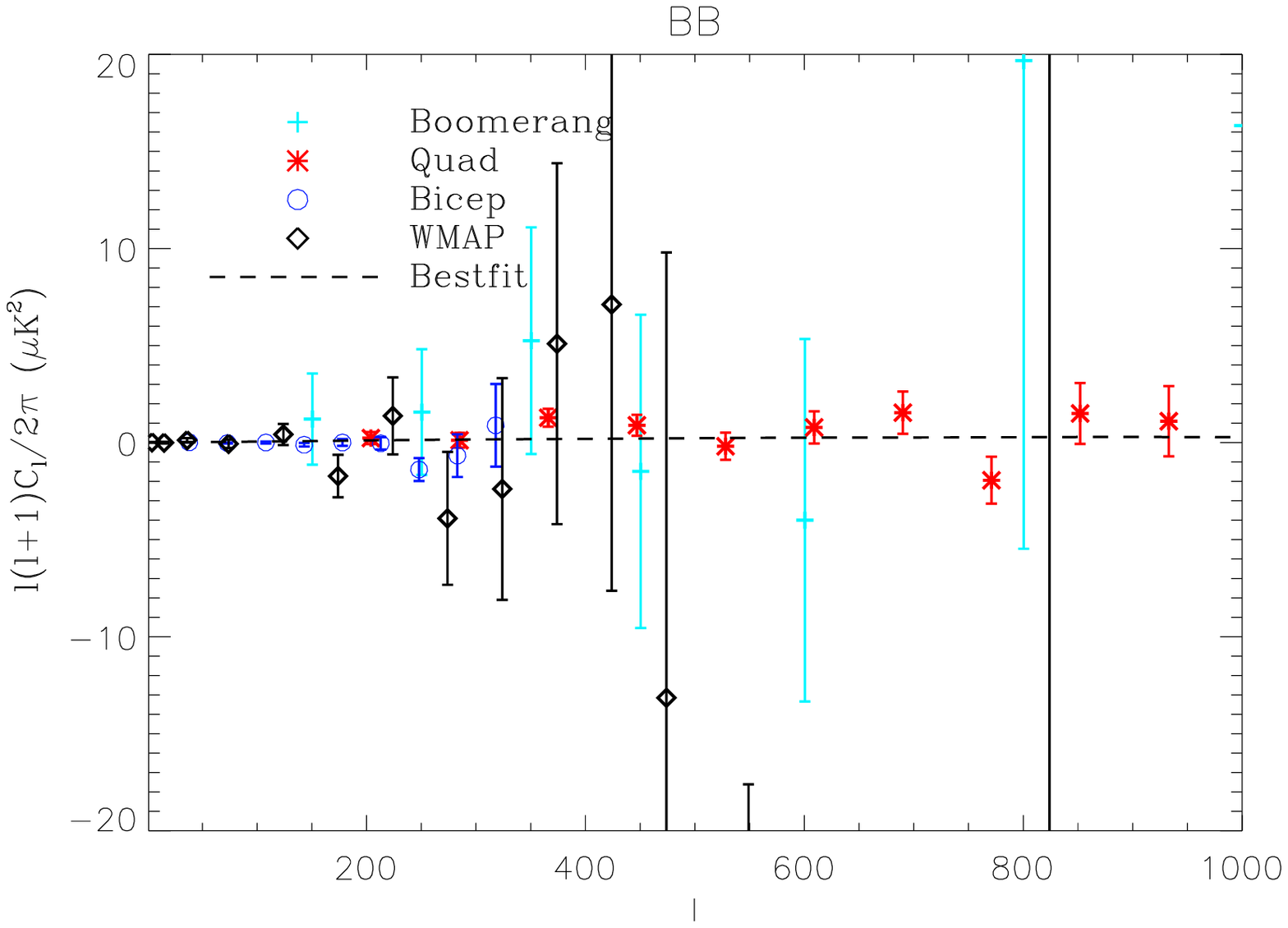}
\end{tabular}
\end{center}
\caption[fig:spettripol] {\label{fig:spettripol}{\bf Left:} Recent
selected measurements of the angular cross-spectrum
Temperature-E-modes-polarization $\langle TE \rangle$. {\bf
Center:} Recent selected measurements of the angular power
spectrum of E-modes-polarization $\langle EE \rangle$. {\bf
Right:} Recent selected measurements (upper limits) of the angular
power spectrum of B-modes-polarization $\langle BB \rangle$. Note
the different vertical scales for the three plots. The dashed line
is the model prediction for the same cosmological parameters best
fitting $\langle TT \rangle$ measurements.}
\end{figure}

Recent measurements of the angular power spectra of CMB
polarization are collected in fig.\ref{fig:spettripol}. The
polarization power spectra measured by these experiments are all
consistent with the forecast from the ``concordance" model best
fitting the \wmap $\langle TT \rangle$ power spectrum. In
addition, they constrain the optical depth to reionization (the
process ionizing the universe when the first massive stars
formed), which is not well constrained by anisotropy measurements
alone (see e.g.\cite{Mort08}).

The WMAP data have sufficient coverage to allow a stacking
analysis and show the irrotational pattern of polarization
pseudovectors around cold and hot spots of the CMB
sky\cite{Koma11}: a clear visual demonstration of the polarization
produced by density perturbations in the early universe.

To date, measurements of the rotational component of the
polarization field $\langle BB \rangle$ resulted in upper limits,
implying a ratio of tensor to scalar fluctuations $r \simlt 0.3$.

\section{OBSERVING THE CMB}
\label{sec:obse}  

The CMB is a diffuse mm-wave source, filling the sky (with a
photon density of $\sim 400 \gamma/cm^3$) and very faint with
respect to radiation produced in the same wavelength range by our
living environment and by the instruments used to measure it (the
telescope, the optical system, the filters, the detector). The
greatest difficulty in measuring the CMB is to reduce the
contamination from other sources.

Measuring the specific brightness of the CMB with the COBE-FIRAS
instrument required cooling cryogenically the spectrometer (a
Martin-Puplett Fourier-Transform Spectrometer) and the bolometric
detectors, and launching it in a 400 km orbit. The first operation
reduced drastically the emission of the instrument and the noise
of the detectors, the second minimized the emission of the earth
atmosphere. The COBE-FIRAS was a null-instrument, comparing the
specific sky brightness, collected by a multi-mode Winston
concentrator\cite{Welf78}, to the brightness of an internal
cryogenic blackbody reference. The output was precisely nulled
(within detector noise) for T$_{ref}$=2.725 K . This implies that
the brightness of the empty sky is a blackbody at the same
temperature, and that the early universe was in thermal
equilibrium. The brightness of a 2.725K blackbody\cite{Fixs09} is
relatively large (compared to the typical noise of mm-wave
detectors), but everything at room temperature emits microwaves in
the same frequency range: the instrument itself, the surrounding
environment, the earth atmosphere. A room-temperature blackbody is
orders of magnitude brighter than the CMB. Low emissivity,
reflective surfaces must be used to shield the instrument, which
needs to be cooled to cryogenic temperatures. Also, to avoid a
very wide dynamic range, a cryogenic reference source should be
used in the comparison. All this drove the design of the
COBE-FIRAS instrument ~\cite{Math99}.

The FIRAS one can be considered a definitive measurement of the
spectrum of the CMB in the mm range: the deviations from a pure
blackbody are less than 0.01\% in the peak region, small enough to
be fully convincing about the thermal nature of the CMB. However,
there are regions of the spectrum where small deviations from a
pure blackbody could be expected.

The ARCADE experiment \cite{Fixs04,Fixs11}, another cryogenic flux
collector working with coherent detectors from a stratospheric
platform, focused on the low frequency end of the spectrum,
looking for cm-wave deviations. In addition to emission from
unresolved extragalactic sources, processes like the reionization
due to the first stars, and particle decays in the early universe,
would heat the diffuse matter, which in turn would cool, injecting
the excess heat in the CMB (see e.g. \cite{Buri95}).

CMB anisotropy and polarization measurements target at much
smaller brightnesses. The specific brightness of CMB anisotropy
(and its polarization) is a modified blackbody
\begin{equation}
\Delta B =  B(\nu, T_{CMB}) \frac{xe^x}{e^x-1} \frac{\Delta T}{T}
\; \; \; \; ; \; \; \; \; x={h \nu \over k T}
\end{equation}
peaking at 220 GHz, and with $\Delta T / T$ of the order of 30
ppm, resulting in very faint brightness differences. However, in
differential measurements common mode signals (coming from the
average CMB but also from the instrument and the environment) can
be rejected with high efficiency.

The focus here shifts on angular resolution (i.e. size of the
telescope), sensitivity (i.e. noise of the detectors and
photon-noise from the radiative background), and mapping speed.

\subsection{Angular Resolution and Sidelobes}

Theory predicts a power spectrum $c_\ell^{**}$ with important
features at degree and sub-degree angular scales (i.e. multipoles
above $\ell = 100$, see top panel of Fig.\ref{fig:spettri}).
Resolving those features requires sub-degree angular resolution.
In the bottom panel of Fig.\ref{fig:spettri} we plot the window
function (i.e. the sensitivity of the instrument to different
multipoles) for Gaussian beams with different $FWHM$: $B_\ell^2 =
e^{-\ell(\ell+1/2)/\sigma^2}$, with $\sigma = FWHM/\sqrt{8 \ln 2}$
\cite{Silk80}.

At the frequency of maximum specific brightness of the CMB (160
GHz), and for a FWHM of 10$^\prime$, the diameter of the entrance
pupil of a diffraction-limited optical system has to be around 0.8
m. However, to reduce the spillover from strong sources in the
sidelobes, it is needed to oversize the entrance pupil (i.e. the
diameter of the collecting mirror/lens) leaving a guard-ring
around the entrance pupil. The aperture stop is placed in a cold
part of the system, effectively apodizing the illumination of
primary light collector. So, at least meter-sized telescopes are
needed to explore the features of the angular power spectrum of
the CMB, while 10m class telescopes are needed to study its finest
details.

Bolometric systems are capable of integrating many radiation
modes, boosting their sensitivity at the cost of a corresponding
increase in the size of the entrance pupil. At lower frequencies,
where the atmosphere is more transparent, the required telescope
size increases by a large factor (for example by a factor $\sim$ 4
at $\sim$ 40 GHz), entering in the realm of large and expensive
telescope structures, including compact interferometric systems.

For all these reasons, CMB telescopes cannot easily be cooled at
cryogenic temperatures to reduce their radiative background,
unless they are operated in space (see below).

The control of telescope sidelobes is also extremely important. If
ground pickup is not properly minimized, the nuisance signal
coming from the earth emission in the far sidelobes can be
comparable or larger than the CMB anisotropy signal. The detector
will receive power from the boresight, pointed to the sky, but
also from all the surrounding sources, weighted by the angular
response $R$ as follows:

\begin{equation} \label{ar}
W(\theta, \phi) = A \int_{4\pi} B(\theta^\prime, \phi^\prime)
R(\theta-\theta^\prime, \phi-\phi^\prime) d\Omega
\end{equation}

where  $B(\theta, \phi)$ is the brightness from direction
$(\theta, \phi)$.

Beyond the main beam (off-axis angles $\theta \gg \lambda / D$)
the envelope of the angular response $R$ for a circular aperture
in diffraction limited conditions scales as $\theta^{-3}$. For a
ground based experiment, where the sky fills the main beam with
solid angle $\Omega_M \ll 1$ sr and the emission from ground fills
a large solid angle in the sidelobes $\Omega_S \sim 2\pi$ sr, the
detected signal can be approximated as

\begin{equation} \label{simple}
W \simeq A \left[ B_{sky} \langle R \rangle_M \Omega_M +
B_{ground} \langle R \rangle_S \Omega_S \right] =
A\left[I_M+I_S\right]
\end{equation}

where $\langle \rangle_{M,S}$ represent the averages of the
angular response $R$ over the main lobe (where $\langle R
\rangle_M \simlt 1$) and over the sidelobes (where $\langle R
\rangle_S \ll 1$). In the case of a 2.725K sky emission and of a
250K ground emission, for example, in order to have $I_S \ll I_M$
we need $\langle R \rangle_S \ll 4 \times 10^{-5}$ for a 10$^o$
FWHM experiment, and
 $\langle R \rangle_S \ll 1 \times
10^{-8}$ for a 10$^\prime$ FWHM experiment. Hence the necessity of
additional shields surrounding the telescope, to increase the
number of diffractions that radiation from the ground must undergo
before reaching the detectors.

The situation is even worse in the case of anisotropy
measurements, where the interesing signal is of a few $\mu K$.
Here a differential instrument is needed, which helps in reducing
the sidelobes contribution to the measured signal. The last
resource is to send the instrument far from the earth, so that the
solid angle occupied by ground emission is $\ll 2\pi$. This is the
case for the WMAP and Planck space missions, devoted to CMB
anisotropy and polarization measurements. They both operate from
the Lagrange point L2 of the sun-earth system, where the solid
angle occupied by the earth is only $2 \times 10^{-4}$ sr. This
relaxes the conditions on sidelobes rejection by a factor $\sim
30000$ with respect to ground-based or balloon-borne experiments.

The telescope and shields configurations are optimized using
numerical methods (see e.g. www.ticra.com), normally based on the
geometrical theory of diffraction \cite{Kell62} to speed-up the
computations (see e.g. \cite{Osul07}).

To reduce the sidelobes, off-axis telescope designs are preferred,
and complemented by extensive ground and sun shields (see e.g.
\cite{Dall88,Beno02,Piac02,Mill02,Cril03,Page03,Mart04,Masi06,Carl11,Swet11}).
In particular, compact test range telescope configurations offer
wide focal planes (allowing the use of large format detector
arrays, see below), with excellent cross-polarization quality
(which is essential for CMB polarization studies) see e.g.
\cite{Grim09,Bude10}.

The actual sidelobes pattern is usually measured with strong
far-field sources (like a Gunn oscillator in the focus of a large
telescope, producing a plane-wave to illuminate the telescope of
the instrument). For space missions, where the operating
environment can be very different from the laboratory conditions,
the sidelobes are measured during the mission, using the Moon or
the Sun (see e.g. \cite{Barn03}).

\subsection{Sensitivity}

\begin{figure}
\begin{center}
\begin{tabular}{c}
   \includegraphics[height=9cm]{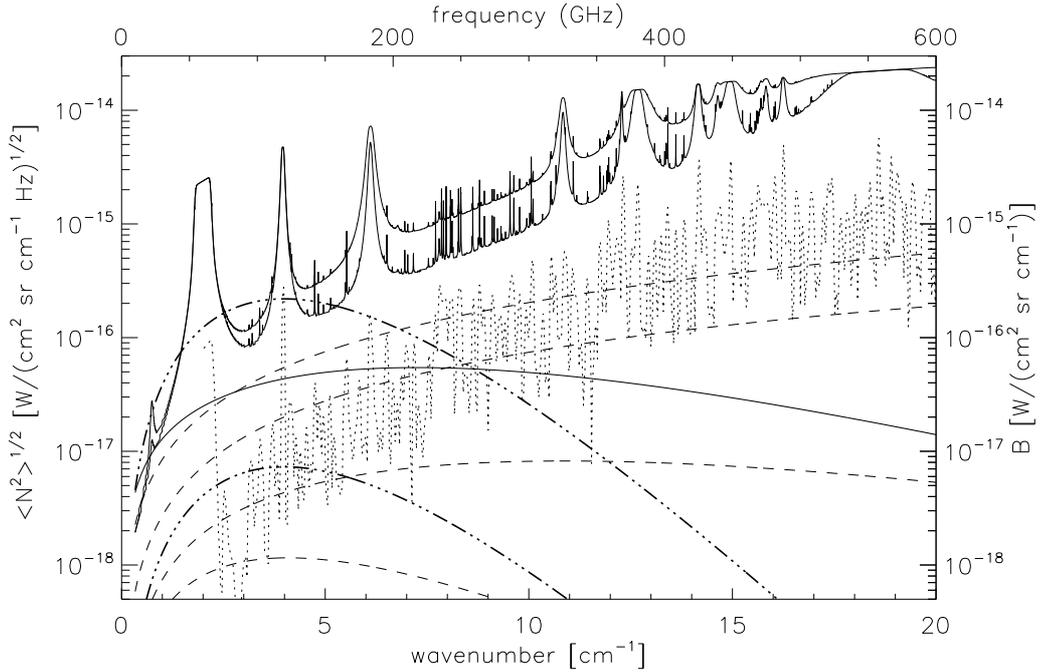}
\end{tabular}
\end{center}
\caption[fig:noise] {\label{fig:noise} Photon noise from the
natural radiative background and from the instrument (left scale)
compared to CMB anisotropy and polarization signals (right scale).
The two top continuous lines represent the noise due to quantum
fluctuations of atmospheric emission, for 2 mm PWV and 0.5 mm PWV,
typical of a high altitude ground based observatory. The dotted
line is the noise due to quantum fluctuations of the emission from
the residual atmosphere, at balloon (41 km) altitude. The lower
thin continuous line is the photon noise of the CMB itself. The
dashed lines represent the noise produced by a low-emissivity (
$\epsilon = 5\times 10^{-3}$ ) optical system at different
temperatures (300K, 40K, 4K, 1.5K from top to bottom). The
dot-dashed lines represent a typical CMB anisotropy brightness
fluctuation (corresponding to $\Delta T_{CMB} = 90 \mu K$, higher
line ) and a typical CMB polarization fluctuation (corresponding
to $\Delta T_{CMB} = 3 \mu K$, lower line). }
\end{figure}

The sensitivity of a detector measuring CMB anisotropy depends on
detector performance (usually quantified by its intrinsic noise
equivalent temperature, $NET_i$, in CMB temperature fluctuation
units ($\mu K_{CMB}$)) and on the noise of the incoming radiative
background, $NET_\gamma$. The latter is computed following
\cite{Lewi47} (but see also \cite{Lama86,Zmui03}): it depends on
the emission of the instrument itself (presence of warm lenses,
mirrors, windows) and on the atmosphere above the operation site
(the telescope can be ground-based, on a stratospheric balloon, or
on a satellite). Operating above the earth atmosphere photon noise
is reduced, and the instrument must be cooled cryogenically to
exploit the optimal environmental conditions. In figure
\ref{fig:noise} we compare the photon noise from the natural
radiative background and from the instrument to the signal to be
detected, for several typical situations. Keep in mind that photon
noise $\langle N^2 \rangle^{1/2}$ in figure \ref{fig:noise} is
given for unit optical bandwidth (1 $cm^{-1}$), unit electrical
bandwidth (1 Hz, roughly corresponding to 1 s of integration), and
for a throughput $A\Omega = 1 cm^2 sr$, and scales as the square
root of these; moreover, in Rayleigh-Jeans conditions, it scales
also as the square root of the emissivity $\epsilon$.

From figure \ref{fig:noise} it is evident that ground-based
observations are limited to low frequencies ($\simlt$ 40 GHz) and
the W and D bands (note, however, that only quantum fluctuations
have been plotted here, while turbulence, winds, instabilities can
increase atmospheric noise significantly). Balloon-borne
telescopes can work with room-temperature telescopes, while to
exploit the low radiative background of space the telescope should
be cooled to at least 40K, and better below 4K if high frequency
measurements are planned.

Quite recently mm-wave bolometers operated below 0.3K have
achieved background limited conditions (i.e. $NET_i \simlt
NET_\gamma$). This is the case of the bolometers of the HFI
instrument\cite{Lama03} aboard of the Planck satellite, where the
telescope is cooled radiatively to 40K\cite{Plan11IV}. For these
detectors $NEP_i \simlt 10^{-17} W/\sqrt{Hz}$ (\cite{Holm08}), and
a cold optical system is required, to exploit their excellent
performance (compare this $NEP_i$ to the photon noise in fig.
\ref{fig:noise}, for a typical throughput $A\Omega \sim \lambda^2
\sim 0.05 cm^2 sr$).

\subsection{Mapping Speed}

Once background-limited conditions are reached, the only way to
improve the performance of a CMB survey is to increase the number
of detectors simultaneously scanning different directions, i.e. to
produce large arrays of mm-wave detectors. This will boost the
mapping speed of the experiment, by a factor of the order of the
number of detectors in the focal plane. The need for large arrays
required an important technology development to achieve fully
automated production of a large number of pixels. This is very
difficult to achieve in the case of coherent detectors, because of
the cost and the power dissipation of each amplifier. In the case
of bolometers and other incoherent detectors (like KIDs and CEBs,
see below), it has been possible to devise pixel architectures
which can be completely produced by photolithography and
micromachining, with low cost and negligible power per pixel.

Bolometers are thermal detectors, absorbing radiation and sensing
the resulting temperature increase. For a review of CMB bolometers
development and operation see e.g. \cite{Rich04,Rich05}. The
development of fully lithographed arrays is the result of a long
process started with the development of the so-called spider-web
bolometer \cite{Maus97}, followed by the polarization sensitive
bolometer (PSB) \cite{Jone03}. Several of these devices were
arranged on the same wafer \cite{Maus00}. Then voltage-biased
Transition Edge superconducting Sensors (TES) were developed (see
e.g. \cite{Irwi95,Lee96,Lee98,Gild99,Gild00} and integrated on the
array wafer (see e.g.
\cite{Benf03,Niem08,Mehl08,Kuo08,Orla10,Pajo10,Schw11,Stan12}). In
parallel to the development of the TES bolometers, a large effort
has been spent in the development of the readout electronics,
which uses SQUIDs to read and multiplex a large number of
detectors with a limited number of wires, thus maintaining the
heat load on the cryostat at manageable levels
\cite{Lant06,Batt08}. These detectors have been installed at large
CMB telescopes (ACT, SPT, APEX ...) with excellent performance,
providing high resolution CMB measurements. Antenna coupling to
the radiation, dual polarization sensitivity, and even spectral
filtering are now also integrated in the TES wafers (see
e.g.\cite{Myer04,Myer05,Myer06,Obri08,Obri09,Obri10,Obri11}),
producing powerful imaging/polarimetry/spectrometry capabilities
in a lightweight block.

In addition to TESs, the quest for large mm-wave cameras for CMB
research drove the development of other non-coherent detection
technologies.

In the MKIDs (microwave kinetic inductance detectors) low energy
photons (like CMB photons, in the meV range) break Cooper pairs in
a superconducting film, changing its surface impedance, and in
particular the kinetic inductance $L_k$. The change is small, but
can be measured using the film as the inductor in a
superconducting resonator, which can have very high merit factor
$Q$, up to $\simeq 10^6$, and thus be very sensitive to the
variations of its components. Many independent MKIDs are arranged
in an array, an shunt the same line, where a comb of frequencies
fitting the resonances of the pixels is carried. CMB photons
absorbed by a given pixel produce a change in the transmission of
a single frequency of the comb. So MKIDs are intrinsically
multiplexable, requiring only two shielded cables to supply and
read hundreds of pixels. The initial KID concept \cite{Day03},
where mm-wave photons are antenna coupled to the resonator, has
evolved in the LEKID (Lumped Elements Kinetic Inductance Detector)
concept \cite{Doyl08,Doyl10}, where the resonator is shaped as an
efficient absorber of mm-waves analogous to bolometer absorbers.
The great advantage of MKIDs with respect to TES is that the
fabrication process is significantly simpler, and also the readout
electronics requires only a wide-band amplifier cryogenically
cooled. Today, MKID arrays are produced in many laboratories (see
e.g.  \cite{Yate08,Malo10,Calv10}) and are starting to be operated
at large telescopes (see e.g. \cite{Monf10,Monf11}).

In a Cold Electron Bolometer (CEB) \cite{Kuzm00} the signal power
collected by an antenna is capacitively coupled to a
tunnel-junction (SIN) sensor and is dissipated in the electrons
which act as a nanoabsorber; it is also removed from the absorber
in the form of hot electrons by the same SIN junctions. This
electron cooling provides strong negative electrothermal feedback,
improving the time-constant, the responsivity and the
NEP\cite{Tara10}. Since the thermistor is the gas of electrons,
confined in a $\sim$ 100 nm junction and thermally insulated from
the rest of the sensor, these detectors should be quite immune to
cosmic-rays hits, an important nuisance for TES and KIDs in space.
Moreover, these detectors promise very good performance in a wide
range of radiative backgrounds, while efficient multiplexing
schemes are still to be developed.

\section{CURRENT TRENDS IN CMB RESEARCH}
\label{sec:current}

We have entered the era of precision observations of the CMB.

\begin{figure}
\begin{center}
\begin{tabular}{c}
\includegraphics[width=5cm]{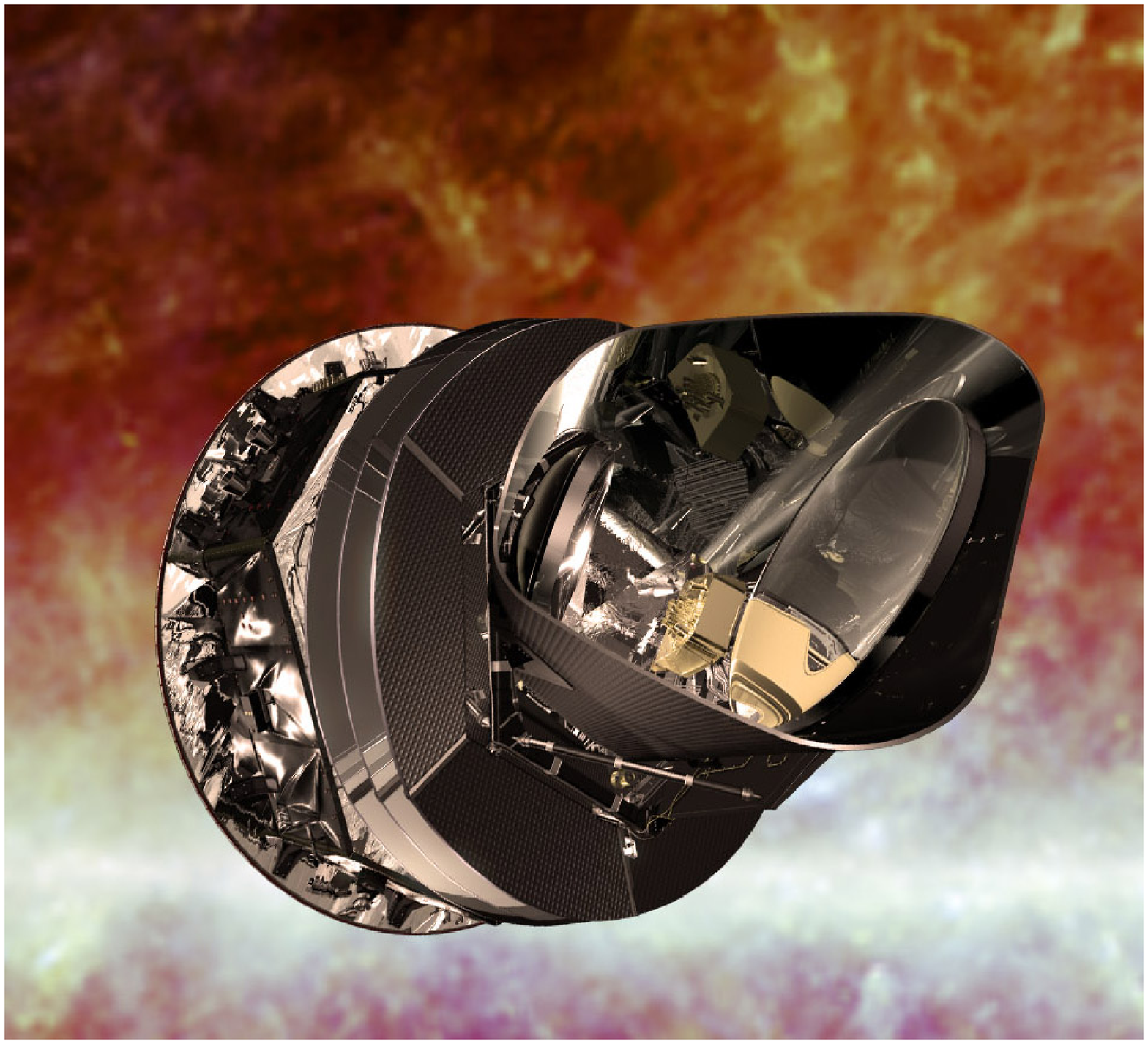}
\includegraphics[width=9.05cm]{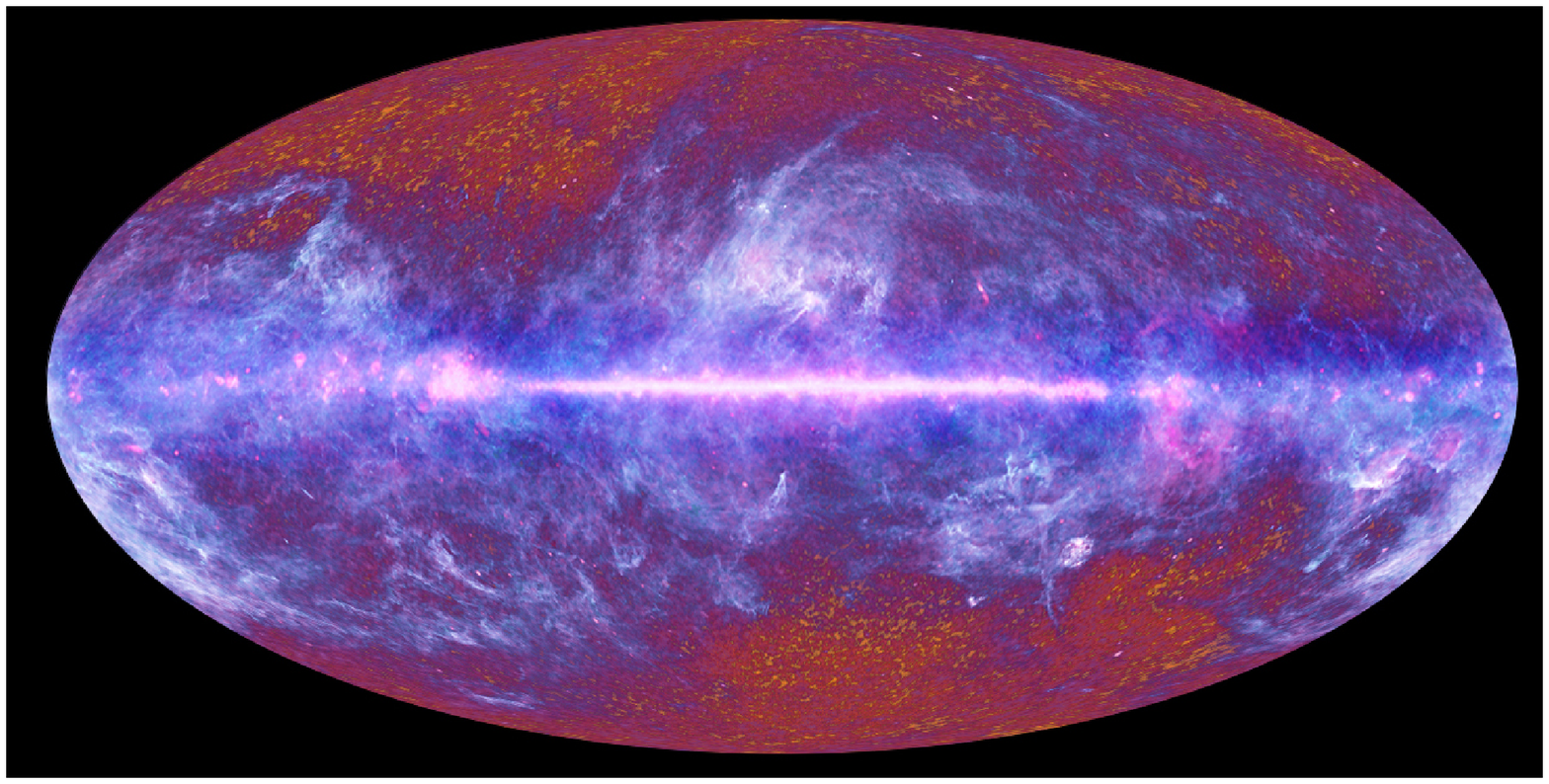}
\end{tabular}
\end{center}
\caption[fig:planck] {\label{fig:planck} {\bf Left:} The Planck
satellite, scanning the sky in the mm/submm from the Lagrangian
point L2 of the sun-earth system. {\bf Right:} Map of the sky in
Galactic coordinates, obtained from the full sky surveys of Planck
at nine frequencies. The maps have been linearly combined in two
ways: the red-yellow palette evident at high galactic latitudes is
the combination maximizing CMB signals. Here the horizon-sized
spots are ubiquitous and dominant at high galactic latitudes. The
blue-white palette is the combination maximizing the local
foreground. The power of wide frequency coverage is evident: from
this map it is possible to monitor the faintest interstellar
clouds at high galactic latitudes. Figure credit: European Space
Agency. }
\end{figure}

Planck \cite{Plan11I} has produced a shallow survey of the whole
sky in nine mm - submm bands (centered at 30, 44, 70, 100, 143,
217, 353, 545, 857 GHz). Taking advantage of the wide frequency
coverage and of the extreme sensitivity of the measurements, it is
possible to separate efficiently the different contributions to
the brightness of the sky along each line of sight (see figure
\ref{fig:planck}).

While it is evident from fig.\ref{fig:planck} that foreground
emission can be important even at high galactic latitudes, it is
also clear that the multifrequency survey of Planck allows to
detect and remove tiny contaminations from thin interstellar
clouds. With the foregrounds under control, Planck is expected to
produce very precise measurements of CMB anisotropy and
polarization in the next data release, early in 2013.

High-resolution anisotropy measurements are now performed mainly
in the direction of clusters of galaxies (SZ effect) and to search
for non-Gaussianity of the CMB. High sensitivity polarization
measurements aim at measuring B-modes from inflation and from
lensing of E-modes. We will outline here a few, selected issues
that we consider relevant for the continuation of these studies.

\subsection{Sunyaev-Zeldovich Effect and spectral anisotropy measurements}

The Sunyaev-Zel\'dovich (SZ) effect\cite{Suny72} is the
energization of CMB photons crossing clusters of galaxies, due to
the inverse Compton effect with the hot intergalactic plasma. The
order of magnitude of the effect can be estimated noticing that
the optical depth for a rich cluster is $\tau \sim n_e
 \sigma_T
\ell \simlt 0.01$ and the fractional energy gain of each
interacting photon is of the order of $kT_e / m_e c^2 \sim 0.01$,
so the fractional CMB temperature change will be $\Delta T / T
\sim \tau kT_e / m_e c^2   \simlt 10^{-4}$: a large signal if
compared to the primordial CMB anisotropy. The SZ effect is thus a
powerful tool for studying the physics of clusters and using them
as cosmological probes (see e.g. \cite{Birk99,Carl02,Reph06}).
Large mm-wave telescopes (\cite{Carl11,Swet11,Schw11}), coupled to
imaging multi-band arrays of bolometers, are now operating in
excellent sites and produce a number of detections and maps of the
SZ effect in selected sky areas, discovering new clusters, and
establishing cluster and cosmological parameters.

From the Planck data an early catalogue of massive clusters
detected via the SZ effect has been extracted\cite{Plan11VIII}.
This consists of 169 known clusters, plus 20 new discoveries,
including exceptional members\cite{Plan11XXVI}. All these
measurements take advantage of the extreme sensitivity of
bolometers, with their excellent performance in the frequency
range 90-600 GHz where the spectral signatures of the SZ effect
lie.

Several components contribute to the signal detected from the line
of sight crossing the cluster: a thermal component due to the
inverse Compton effect; a Doppler component, caused by the
collective motion of the cluster with respect to the CMB
restframe; a non-thermal component caused by a non-thermal
population of electrons, produced by e.g. the AGNs present in the
cluster, relativistic plasma in cluster cavities, shock
acceleration; the intrinsic anisotropy of the CMB; the emission of
dust, free-free and synchrotron in our Galaxy and in the galaxies
of the cluster. Since the spectrum of thermal SZ significantly
departs from the spectra of the foreground and background
components, multi-frequency SZ measurements allow the estimation
of several physical parameters of the cluster, provided there are
more observation bands than parameters to be determined, or some
of the contributions are known to be negligible. In \cite{debe12}
we have analyzed how different experimental configurations perform
in this particular components separation exercise. Ground-based
few-band photometers cannot provide enough information to separate
all physical components, because atmospheric noise limits the
number of useful independent bands. These instruments need
external information (optical, X-ray, far-IR, etc.) to produce
mainly measurements of the optical depth of the thermal SZ (see
e.g.
\cite{Hinc10,Marr11,Brod10,Hand11,Sehg11,Fole11,Stor11,Will11}).

\begin{figure}
\begin{center}
\begin{tabular}{c}
\includegraphics[width=4.8cm, height=6cm]{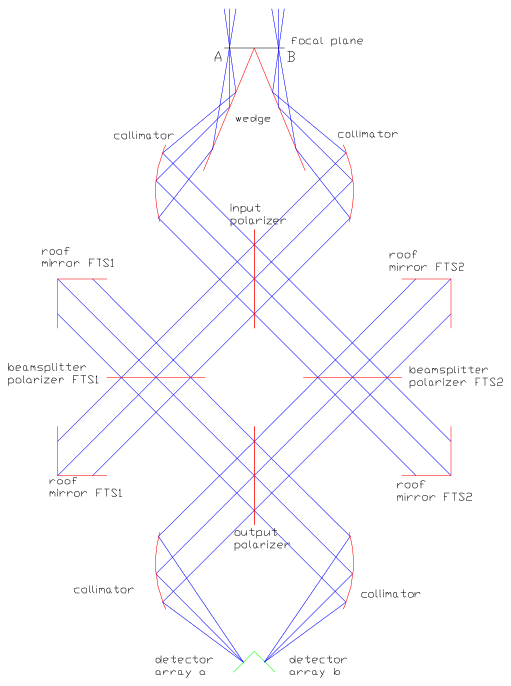}
\includegraphics[width=5.8cm, height=6cm]{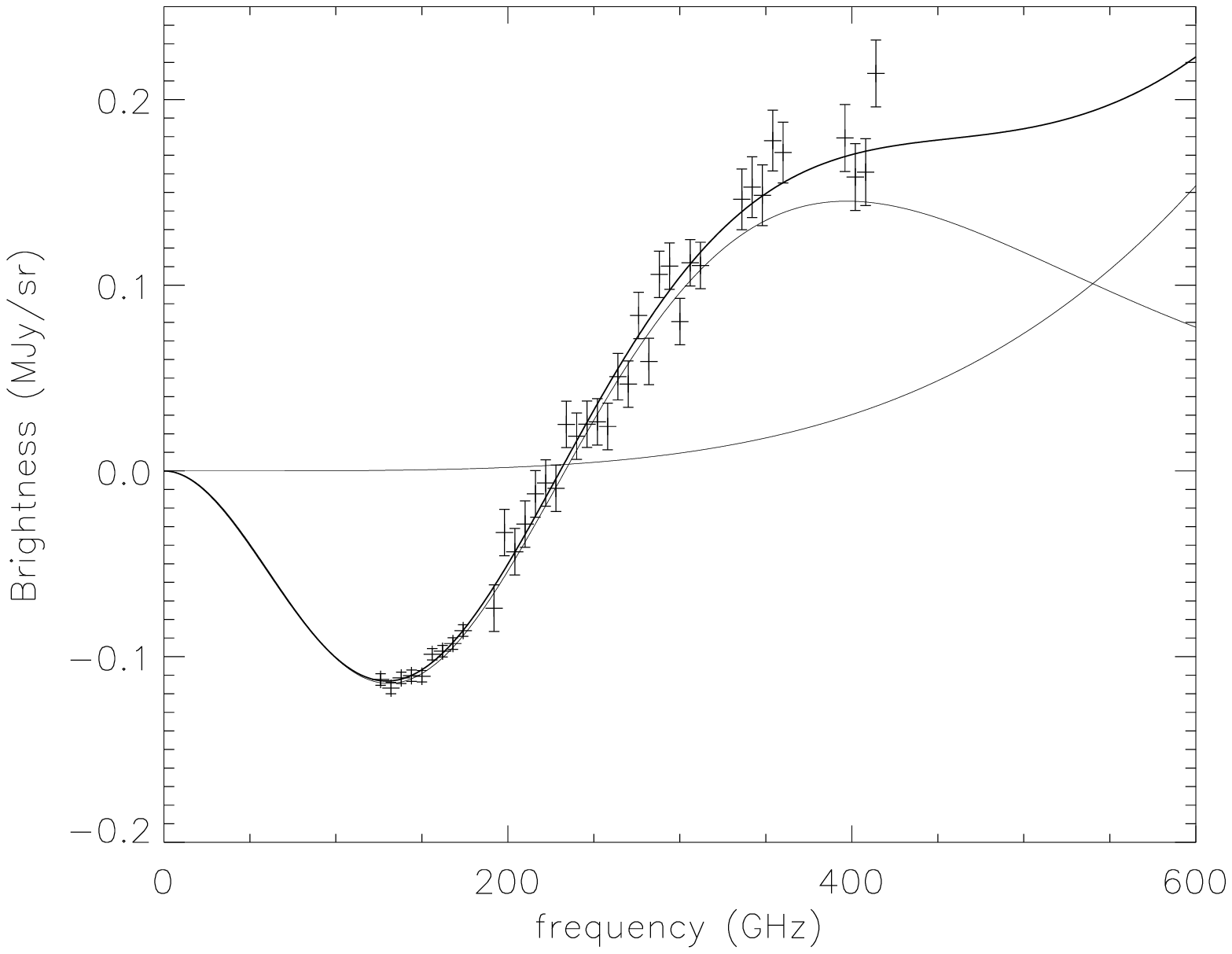}
\includegraphics[width=5.8cm, height=6cm]{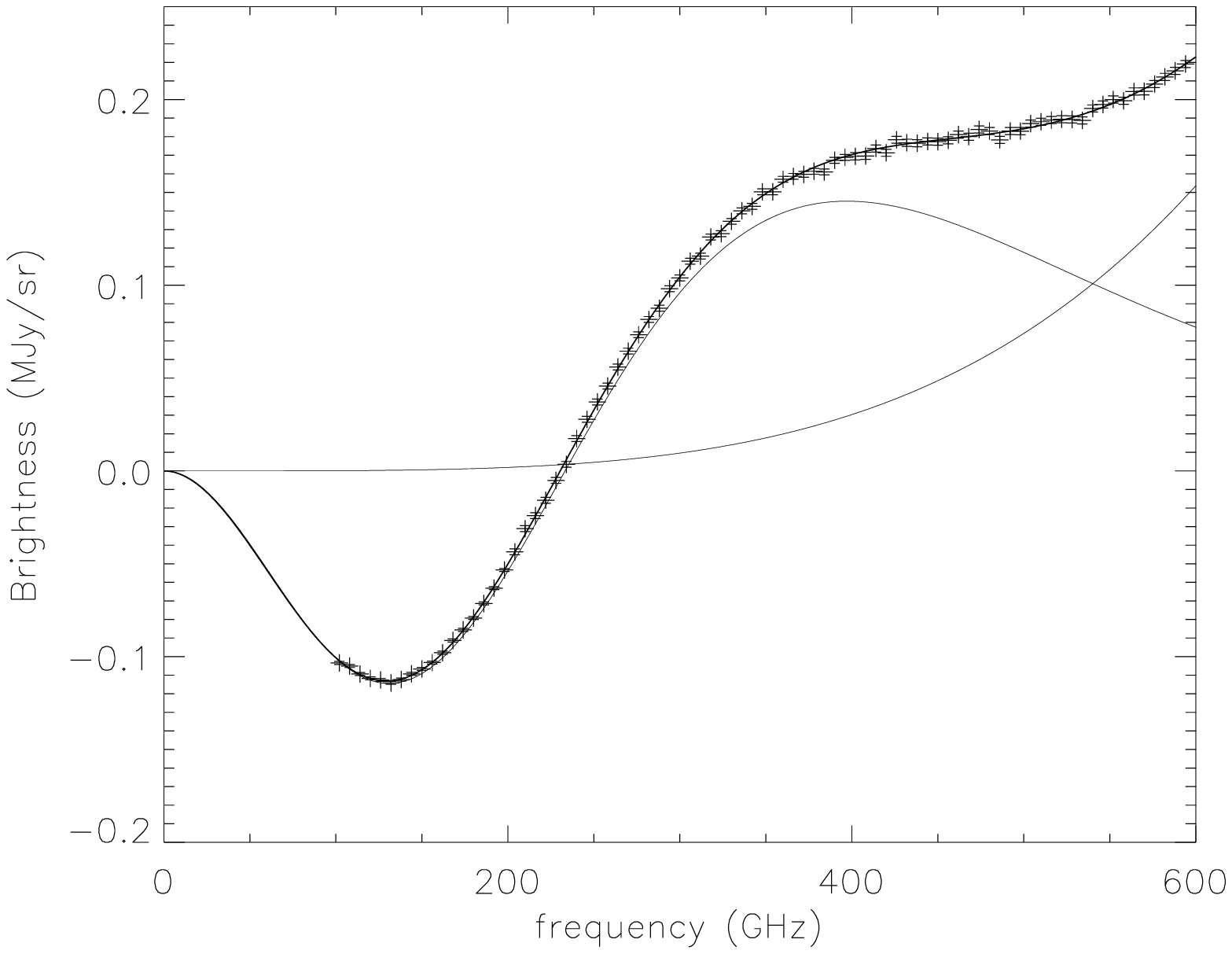}
\end{tabular}
\end{center}
\caption[fig:SZ] {\label{fig:SZ} {\bf Left} : Block-diagram of a
differential FTS. This very symmetrical configuration reduces
instrumental offsets and doubles the efficiency with respect to
the standard MPI FTS\cite{Mart70}. {\bf Center} : Simulated
observations of a rich cluster of galaxies with a warm
differential FTS aboard of a stratospheric balloon, like OLIMPO.
{\bf Right} : Same for a differential FTS aboard of a satellite in
L2, with a large (10m) cold (4K) telescope, like Millimetron. In
both cases, 3 hours of observation are assumed, and bolometer
performance limited only by the radiative background. The two
continuous lines represent the SZ effect from the plasma in the
cluster and differential emission of interstellar dust, the two
main components of the measured brightness.}
\end{figure}

Future space-based spectrometers can cover the full range of
interesting frequencies and offer much more information. A
cryogenic differential imaging Fourier Transform Spectrometer
(FTS) in the focal plane of a space mission with a cold telescope,
like Millimetron\cite{Wild09}, would be a powerful experiment,
measuring accurately all the parameters of a cluster. The FIRAS
experiment has demonstrated the power of these large-throughput,
wide frequency coverage instruments, which are intrinsically
differential. In that case one of the two input ports collected
radiation from the sky, while the other port was illuminated by an
internal blackbody: the perfect nulling of the measured difference
spectrum demonstrated accurately the blackbody nature of the
cosmic microwave background. In this implementation, the two input
ports of the instrument collect radiation from two contiguous
regions of the focal plane of the same telescope (see left panel
of fig.\ref{fig:SZ}). In this way only the anisotropic component
of the brightness distribution produces a measurable signal, while
the common-mode signals from the instrument, the telescope, and
the CMB itself are efficiently rejected. The FTS is sensitive to a
wide frequency band (say 70 - 1000 GHz for SZ studies), so photon
noise is the limiting factor. In fig.\ref{fig:SZ} we compare what
can be achieved with a warm system on a stratospheric
balloon\cite{Masi08,Conv10,Schi12} (center panel) to what can be
ultimately achieved with a cold system in deep space\cite{debe12}
(right panel). In both cases important improvements with respect
to the state-of-the-art determination of cluster parameters are
expected. The intermediate case of a cold spectrometer coupled to
a warm telescope in a Molniya orbit has been studied in
\cite{debe10} .

Other important scientific targets of these instruments are the
measurement of the $C^+$ and CO lines, in the redshift desert and
beyond, for a large number of galaxies, and spectral observations
of a number of processes in the early universe and in the
recombination and reionization eras (see e.g.
\cite{debe93,Basu04,Dubr08,debe10,Kogu11,Gong12}).

\subsection{B-modes of CMB polarization}

Measuring the tiny B-modes signal is a formidable experimental
challenge. For this reason it is very important that independent
teams develop advanced experiments, using different techniques and
methods. Only independent consistent detections will provide
convincing evidence for the existence of B-modes.

The mainstream in this field is the use of large arrays of
single-mode bolometric polarimeters, using a polarization
modulator in the optical path (as close as possible to the input
port of the instrument, to avoid modulating instrumental
polarization) to modulate only the polarized part of the incoming
signal. The throughput of the telescope has to be very large, of
the order of $\sum_{i=1}^n N_i\lambda_i^2/F_i$, where the number
of detectors $N_i$ in each band $i$ is of the order of 10$^3$, and
the filling factor of the focal plane is $F_i \simlt 1$.

The removal of the polarized foreground (mainly produced by the
interstellar medium) is a matter of the utmost importance. It has
been analyzed in great detail, most recently in the framework of
the Planck mission (see e.g. \cite{Leac08,Ricc10,Armi11} and
references therein) and of future missions devoted to CMB
polarization (see e.g. \cite{Erra12}). The solution is to carry
out surveys with wide frequency coverage, from tens of GHz (to
survey strongly polarized synchrotron emission) to several
hundreds of GHz (to survey polarized emission from interstellar
dust). For this reason a number of bands $n \sim 10$ is required
to separate the foreground components from the primordial CMB
signal. Accommodating all the bands in the focal plane of the
telescope exacerbates the large throughput problem for these
systems, also because only the center region of the focal plane
has optimal polarization efficiency and beam-symmetry properties.
Multichroic pixels including bolometric detectors in 3 or more
bands under the same microlens have been developed \cite{Obri11},
allowing a very efficient use of focal plane space. This approach
has been proposed for the LiteBIRD satellite \cite{Hazu11}.

In the case of incoherent detectors, intrinsically insensitive to
the polarization status of the incoming power, the classic Stokes
polarimeter requires a half-wave plate retarder plus a polarizer.
If the HWP is rotated with a rotation rate ${\dot \theta}$, the
linearly polarized part of the incoming signal is modulated at
$4{\dot \theta}$, while the unpolarized and the circular
polarization components are not modulated. Wide-band retarders can
be obtained in transmission using a sandwich of birefringent
crystals (see e.g. \cite{Hana05,John07,Savi06,Pisa06,Brya10}) or
suitable meta-materials assembled with metal meshes \cite{Zhan11}.
In reflection, a rotation mirror / polarizer combination
\cite{Siri04} can be used, or a translating polarizer / mirror
assembly (Variable Delay Modulator \cite{Krej08}), or a
translating circular polarizer / mirror combination (Transational
Polarization Rotator \cite{Chus12}). The main issues with these
modulators is the equalization of the transmission (reflection)
for the two orthogonal polarizations (any mismatch, even at a
level of 1$\%$, will produce a comparatively very large $2{\dot
\theta}$ signal) and the need to cool at cryogenic temperatures
the modulator, to reduce its (polarized) emission (see
\cite{Sala11} for a discussion).

Reaching satisfactory performance over a wide frequency band and a
wide throughput is problematic. In the case of the dielectric HWP,
a sandwich of differently oriented plates is required, following
the Pancharatman \cite{Panc55} recipe. This approach is suitable
for accurate measurements of CMB polarization in the range 120-450
GHz\cite{Bao12}. However, it is currently impossible to obtain
large-diameter ($\simgt 30 cm$) slabs of sapphire (or any other
birefringent crystal suitable for mm wavelengths), so their use is
limited to medium throughput systems. Using metal meshes might
solve the problem, but requires a careful equalization of the
conductivity of the meshes. In the case of the mirror/polarizer
combination, which can be produced in very large sizes, the
operative band is restricted to $\sim 20\%$ of the center
frequency. It is possible, however, to operate the modulator at
multiples of a fundamental frequency, with decreasing fractional
bandwidth, as proposed in \cite{Core11}.

A Martin-Puplett Fourier Transform Spectrometer\cite{Mart70}, with
the two input ports $A$ and $B$ (fig.\ref{fig:SZ}) co-aligned to
look at the same sky patch, becomes a polarimeter. In fact, it
produces at the two output ports $a$ and $b$, with polarization
$x$ and $y$, the following 4 signals, which can be detected by 4
independent detectors: $ [I_{a,x}(z)-\langle I_{a,x} \rangle]
\propto \int (E^2_{B,x}(\sigma)-E^2_{A,y}(\sigma))cos(4\pi \sigma
z) d\sigma $ ; $
 [ I_{a,y}(z)-\langle I_{a,y} \rangle ] \propto \int
(E^2_{B,y}(\sigma)-E^2_{A,x}(\sigma))cos(4\pi \sigma z) d\sigma $
; $
 [I_{b,x}(z)-\langle I_{b,x} \rangle ] \propto \int
(E^2_{A,x}(\sigma)-E^2_{B,y}(\sigma))cos(4\pi \sigma z) d\sigma $
; $ [I_{b,y}(z)-\langle I_{b,y} \rangle ]\propto \int
(E^2_{A,y}(\sigma)-E^2_{B,x}(\sigma))cos(4\pi \sigma z) d\sigma$ ,
where $\sigma$ is the wavenumber and $z$ is the position of the
moving mirror. Summing and subtracting the Fourier-transformed
signals from detectors couples it is possible to estimate the $\sl
frequency$ $\sl spectra$ of the Stokes parameters of the incoming
radiation. This is the principle of operation of the proposed
PIXIE experiment\cite{Kogu11}, a space-based large-throughput
spectro-polarimeter covering the frequency range 30-6000 GHz. The
optical axis of the spectro-polarimeter is aligned to the spin
axis of the satellite, so that any polarization signal becomes
spin-synchronous. In this configuration, the specifications for
beam ellipticity, and beam, gain and polarization mismatch for the
four detectors are very stringent. These could be relaxed with the
use of a rotating achromatic HWP at the entrance of the system,
but it is currently impossible to fabricate a high-efficiency
highly-balanced HWP over such a wide frequency range.

There is a long list of potential systematic effects in Stokes
polarimeters (see e.g. \cite{Sala10}), and the requirements for a
clean detection of B-modes are extremely stringent (see e.g. table
6.1 in \cite{Bock06}). A few examples: tens of mK signals at $2
{\dot \theta}$ are produced by the unpolarized 2.7K background,
modulated by $\sim 1 \%$ efficiency mismatch between the ordinary
and extraordinary rays in the waveplate. The emission of a
mismatched HWP also produces tens of mK signals at $2{\dot
\theta}$, unless its temperature is below 2K. These signals
challenge the dynamic range of the detector, which is optimized
for measuring CMB polarization signals $\sim 10^4$ times smaller.
Any non-linearity in the detector can convert part of this $2{\dot
\theta}$ signal into a $4{\dot \theta}$ signal, producing a large
offset in the polarization measurement. If part of the emission of
the polarizer is reflected back by the waveplate, it is modulated
at $4{\dot \theta}$, contributing with additional $\sim$ few $\mu
K$ signals to the offset. A possible solution to this problem is
the step and integrate strategy (see e.g. \cite{debe13}). At
variance with the continuous rotation strategy, here the HWP is
kept steady during sky scans, and angular steps are performed at
the turnarounds. All the systematic effects generated internally
to the instrument produce a constant offset during each scan,
which can be removed, while the sky polarization is modulated at
the (very low) frequency of the repetition of the scans. In
addition to these effect, other noticeable sources of systematic
problems are the ellipticity of the main beam ($<10^{-4}$), the
level of its polarized sidelobes ($<10^{-6}$), the instrumental
polarization ($<10^{-4}$), the relative gain calibration
($<10^{-5}$): all these convert unpolarized brightness
fluctuations into apparent B-modes signals; an error in the main
polarimeter axis angle ($<0.2^o$) and the cross-polar response
($<3\times 10^{-3}$) convert E-modes into apparent B-modes;
moreover, the relative pointing of differenced observation
directions must be $<0.1$ arcsec to avoid conversion of brightness
fluctuations into apparent B-mode signals. Pathfinder experiments
are the best way to find and test the best mitigation methods for
all these subtle systematic effects. Current attempts exploit
different techniques, ranging from ground-based coherent
polarimeters, like QUIET\cite{Bisc11}, to ground-based bolometer
arrays with HWP, like POLARBEAR\cite{Arno10}, to ground-based
bolometric interferometers, like QUBIC\cite{Batt11}, to
stratospheric balloons like SPIDER\cite{Fili10},
EBEX\cite{Reic10}, and LSPE\cite{LSPE12}. Using completely
independent techniques, these experiments provide a powerful test
set for any detection of B-modes in the CMB, in view of a
post-Planck next generation space mission for the CMB.

\section{CONCLUSIONS}
\label{sec:conclusions}

The future of CMB studies is bright. A large community has grown
around the success of CMB missions, producing large amounts of
excellent data. The experiments have drifted from a situation
where sensitivity was the issue to a situation where control of
systematic effects is the main problem. So we are facing very
difficult challenges, with the ambition of understanding the most
distant phenomena happening in our universe, analyzing tiny
signals embedded in an overwhelming noisy background. When we
approached CMB research for the first time, in 1980, measuring the
intrinsic anisotropy of the CMB was considered almost
science-fiction. Today CMB anisotropy is measured in a single pass
with scanning telescopes using large arrays of bolometers. This
experience makes us confident that much more is coming in this
field, with the enthusiastic contribution of young researchers and
the cross-fertilization between cosmologists, astrophysicists,
solid-state / detector physicists, optics experts.

\acknowledgments     

This work has been supported by the Italian Space Agency contract
I/043/11/0 "A differential spectrometer for Millimetron - phase A"
and I/022/11/0 "Large Scale Polarization Explorer", and by the
Italian MIUR PRIN-2009 project "mm and sub-mm spectroscopy for
high resolution studies of primeval galaxies and clusters of
galaxies". We thank dr. Luca Pagano for assembling the collection
of CMB measurements in fig.\ref{fig:spettri} and
fig.\ref{fig:spettripol}.


\bibliography{PDB}   
\bibliographystyle{spiebib}   

\end{document}